\documentclass[final,5p,times,twocolumn]{elsarticle}

\usepackage{graphics}

\usepackage{epsfig}

\usepackage{amssymb}

\journal{Radiation Measurements}

\begin{document}

\begin{frontmatter}

\title{Fragmentation cross sections of Fe$^{26+}$, 
Si$^{14+}$ and C$^{6+}$ ions of $0.3 \div 10$ A GeV \\
on CR39, polyethylene and aluminum targets}

\author[label1,label3]{M. Giorgini}

\author{for the BIOSHIELD-B44 and BINFRA-15B517 experiments\corref{cor1}}

 \address[label1]{Dept. of Physics, University of Bologna, I-40127 Bologna, 
 Italy}
 \address[label3]{INFN-Bologna, I-40127 Bologna, Italy \par~\par

{\normalsize \em Talk given at the $24^{th}$ International Conference on
Nuclear Tracks in Solids, \\ Bologna, Italy, 1-5 September 2008.}}

\cortext[cor1]{The {\bf BIOSHIELD-B44} and {\bf BINFRA-15B517} 
 experiments: S. Cecchini, T. Chiarusi, G. Giacomelli, M. Giorgini,
 A. Kumar, G. Mandrioli, S. Manzoor, A.R. Margiotta,
 E. Medinaceli, L. Patrizii, V. Popa, I.E. Qureshi, Z. Sahnoun, G. Sirri,
 M. Spurio and V. Togo}

\begin{abstract}

New measurements of the total and 
partial fragmentation cross sections in the energy range $0.3 \div 10$ A GeV 
of Fe$^{26+}$, Si$^{14+}$ and C$^{6+}$ beams on polyethylene, CR39 and 
aluminum targets are presented. The exposures were made at Brookhaven National 
Laboratory (BNL), USA, and Heavy Ion Medical Accelerator in 
Chiba (HIMAC), Japan. The 
CR39 nuclear track detectors were used to identify the incident and survived 
beams and their fragments. The total fragmentation cross sections for all 
targets are almost energy independent while they depend on the target 
mass. The measured partial fragmentation cross sections are also discussed.

\end{abstract}

\end{frontmatter}

\section{Introduction}
\label{intro}

The interaction and propagation of intermediate and high energy heavy ions in 
matter is a subject of interest in the fields of astrophysics, radio-biology 
and radiation protection (Chen 1994). 

Recently the attention was focused on nucleus-nucleus interactions at
lab energies $\leq$10 AGeV, which are important for radiotherapy and 
radioprotection purposes. The FLUKA (Fass\`o 2003) MonteCarlo code was 
fused with the RQMD (Relativistic Quantum Molecular Dynamics) code
(Garzelli 2006). It is important that the simulated data be continuously 
improved and  
confirmed by new experimental data, such as those presented here.

The availability of ion beams at the BNL (USA) and at 
the HIMAC (Japan) facilities made possible to investigate the projectile 
fragmentation on different targets and for different projectile energies. 

\begin{figure*}[ht]
\begin{center}
{\centering\resizebox*{!}{6cm}{\includegraphics{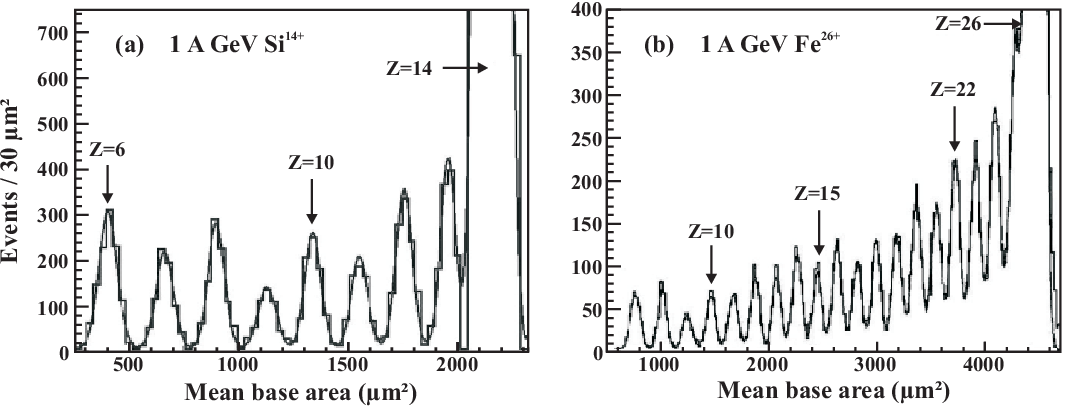}}\par}
\begin{quote}
\caption{\small Distributions of the average base areas for tracks present
in at least 2 out of 3 measured CR39 sheets located after the CH$_2$ 
target. The data concern (a) 1 A GeV Si$^{14+}$ and (b) 1 A GeV Fe$^{26+}$ 
ions. Each peak has a gaussian shape with $\sigma \sim 0.2e$.} 
\label{fig:2}
\end{quote}
\end{center}
\end{figure*}

The present study is focused on Fe$^{26+}$, Si$^{14+}$ and C$^{6+}$ ion 
interactions in CH$_2$, CR39 
$(C_{12}H_{18}O_7)_n$ and Al targets. We used CR39 detectors, which 
are sensitive for a wide range of charges down to $Z = 6e$ in the low 
velocity and in the relativistic regions (Balestra 2007; 
Cecchini 1993; Cecchini 2001; Cecchini 2002; Dekhissi 2000; Giacomelli 
1998). NTD's have been used to 
search for exotic particles like Magnetic Monopoles, Strange Quark Matter and
Q-balls (Ambrosio 2002; Balestra 2008; Cecchini 2008; Medinaceli 2008; 
Sahnoun 2008), to 
study cosmic ray composition (Chiarusi 2005) and for environmental 
studies (Manzoor 2007).

\section{Experimental procedure}
Stacks composed of several CR39 NTD's, of size $11.5 \times 11.5$ cm$^2$, and 
of different targets were exposed to 0.3, 1, 3, 5 and 10 A GeV Fe$^{26+}$, 1, 
3, 5 A GeV Si$^{14+}$ ions at the BNL Alternating Gradient Synchrotron (AGS)
and NASA Space Radiation Laboratory (NSRL). 
Our experiment, named BIOSHIELD-B44, was included in the time slots allocated 
to NASA for a program of space radiation research.
For these low dose experiments, an ionization chamber was used as monitor. It 
was calibrated with a $1 \times 1$ mm$^2$ scintillation counter placed at 
the center of the beam. The beam density was monitored with a pixel counter 
(Lowenstein, D.I. and Rusek, A., 2007) and later checked with our nuclear 
track detectors.  

The exposures to 0.41 A GeV 
Fe$^{26+}$, 0.29 A GeV C$^{6+}$ ions at HIMAC were performed in collaboration 
with colleagues from Naples and Japan. The experiment was named 
BINFRA-15B517. In this 
case, the beam was monitored with a photographic sheet, as explained in
details in Durante 2002.    

For our fragmentation studies, we used three and four CR39 sheets, 
 $\sim$0.7 mm thick, placed before and after the target, respectively. The 
exposures were done at normal incidence, with a
density of $\sim$2000 ions/cm$^2$. 

After exposures the CR39 foils were etched in 6N NaOH aqueous solution at 
70 $^\circ$C for 30 h (in two steps 15h+15h) in a thermostatic water bath 
with constant stirring of the solution. After etching, the beam ions and their 
fragments manifest in the CR39 NTD's as etch pit cones on both sides of each 
detector foil.

The total charge changing cross sections were determined from the survival 
fraction of ions using the following relation
 \begin{equation}
 \sigma_{tot} = \frac {A_T \ln (N_{in} / N_{out})}{\rho~ t ~N_{Av}}
\end{equation}
where $A_T$ is the nuclear mass of the target (average nuclear mass in case 
of polymers: $A_{CH2} = 4.7,~ A_{CR39} = 7.4$); $N_{in}$ and $N_{out}$ are the 
numbers of incident ions before and after the target, respectively; $\rho$ 
(g/cm$^3$) is the target density; $t$ (cm) is the thickness of the target and 
$N_{Av}$ is Avogadro number.\par

Systematic uncertainties in $\sigma_{tot}$ were estimated to be smaller 
than $10\%$: contributions arise from the measurements of the density 
and thickness of the targets, from the separation of the beam peak from the 
 fragments immediately close by (Fig. \ref{fig:2}), from 
fragmentation in the 
CR39 foils and from the tracking procedure.  \par

\begin{figure*}[ht]
\centering
 {\centering\resizebox*{!}{6.1cm}{\includegraphics{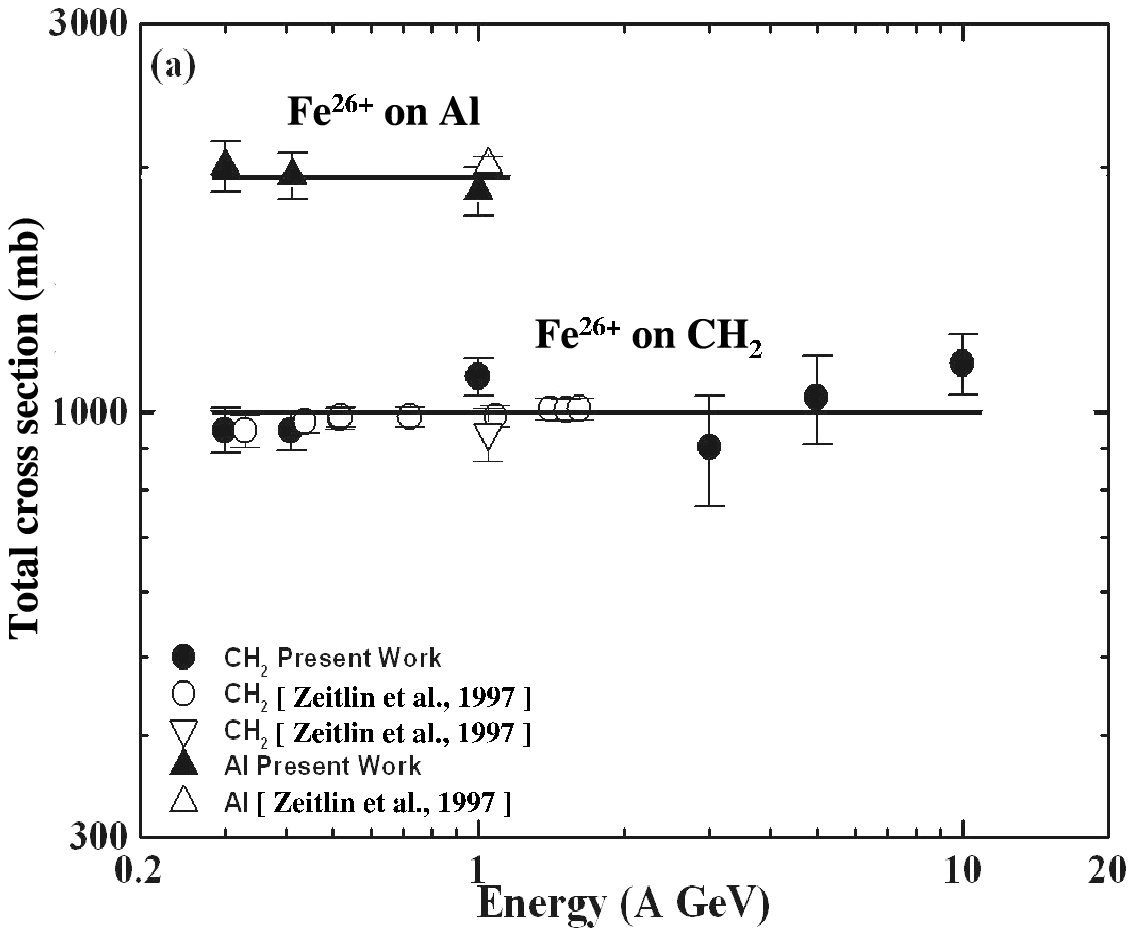}}}
  \hspace{0.8cm}
 {\centering\resizebox*{!}{6.1cm}{\includegraphics{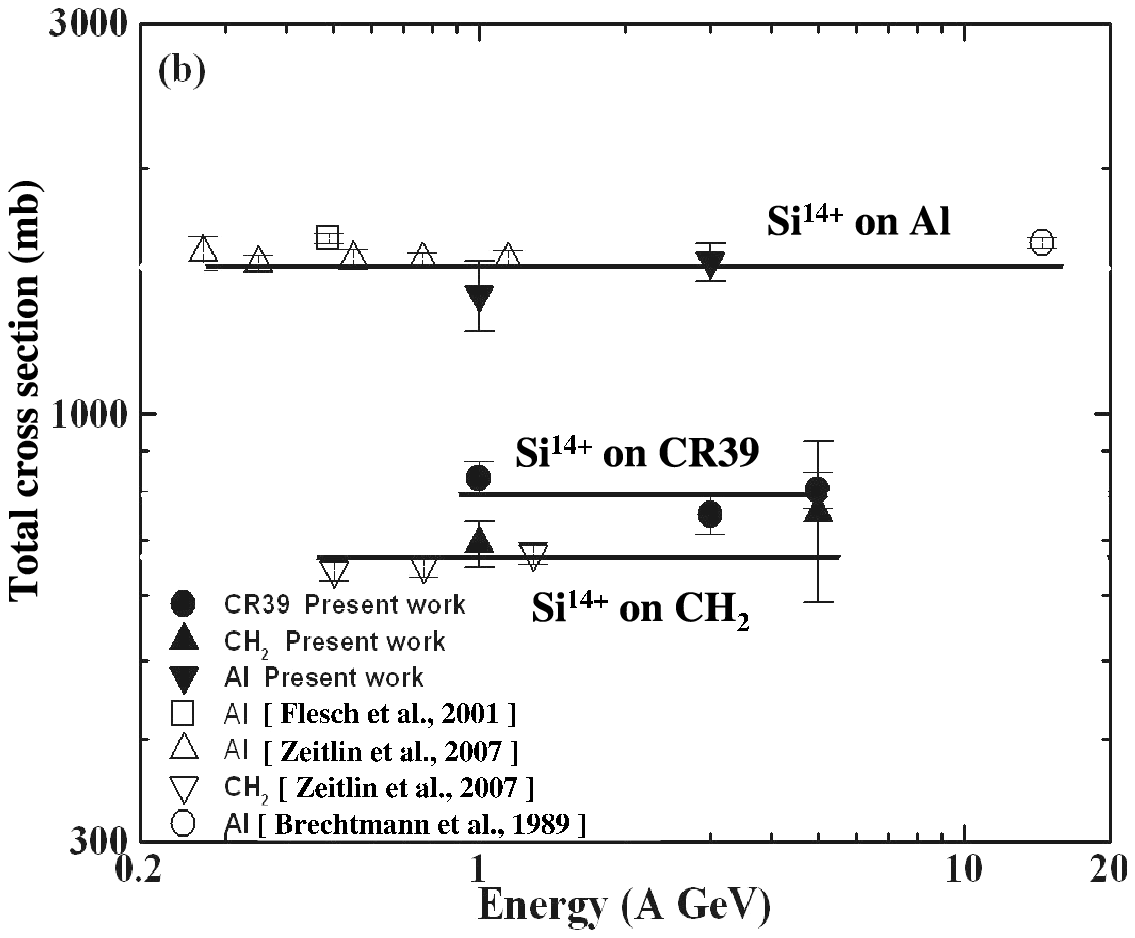}}\par}
\begin{quote}
\caption{\small Total fragmentation cross sections for (a) Fe$^{26+}$ ions of 
  different energies in CH$_2$ and Al targets and (b) for Si$^{14+}$ ions 
in CH$_2$, CR39 and Al targets. The measured cross sections from the cited 
refs. and the predictions from Eq. \ref{eq:2} are shown for 
comparison.} 
\label{fig:3}
\end{quote}
 \end{figure*}

The base areas of the etch-pit cones (``tracks''), their eccentricity and 
central brightness were measured with an automatic image analyzer system 
(Noll 1988) which also provides their absolute coordinates. A 
tracking procedure 
was used to reconstruct the path of beam ions through the front faces of  
the detector upstream (with respect to the target) foils; a similar 
tracking procedure was performed through 
the three measured front faces of downstream CR39 detectors. The average 
track base area was computed for each reconstructed ion path by requiring 
the existence of signals in at least two out of three sheets of the detectors.
In Fig. \ref{fig:2}a,b the average base area distributions for 1 A GeV 
Si$^{14+}$ and 1 A GeV Fe$^{26+}$ beam ions and their fragments 
after the CH$_2$ targets are shown.

\section{Total fragmentation cross sections} 

The numbers of incident and survived beam ions were determined considering 
the mean area distributions of the beam peaks before and after the target 
and evaluating the integral of the gaussian fit of the beam peaks.

The measured total charge changing cross sections are given in the $4^{th}$ 
column of Table \ref{table:1}. Fig. \ref{fig:3}a shows the total cross 
sections of Fe$^{26+}$ projectiles at various beam energies on the CH$_2$ 
and Al targets. Our results for 
Si$^{14+}$ and C$^{6+}$ projectiles are given in Table \ref{table:2} and are 
plotted vs energy in Fig. \ref{fig:3}b. 

The total cross sections are almost energy independent, in agreement 
with the data from other authors (Brechtmann 1988; Brechtmann 1989; Flesch 
 2001; Zeitlin 1997; Zeitlin 2007).

\begin{table}
\begin{center}
{\small
\begin{tabular}
{|c|c|c|c|}\hline
{\bf Energy } & {\bf Target} & {\bf {\boldmath A$_T$}} & {\bf {\boldmath $\sigma_{tot}$} (mb)} \\
 {\bf (A GeV)} & & & \\ \hline 
 10 & CH$_2 $& 4.7 & 1147 $\pm$ 97 \\ \hline
 10 & CR39 & 7.4 & 1105 $\pm$ 360\\ \hline
 5 & CH$_2 $ & 4.7 & 1041 $\pm$ 130 \\ \hline
 5 & CR39 & 7.4 & 1170 $\pm$ 470 \\ \hline
 3 & CH$_2 $ & 4.7 & 904 $\pm$ 140 \\ \hline
 3 & CR39 & 7.4 & 1166 $\pm$ 67 \\ \hline
 1 & CH$_2 $ & 4.7 & 1105 $\pm$ 60\\ \hline
 1 & CR39 & 7.4 & 1113 $\pm$ 176 \\ \hline
1 & Al & 27 & 1870 $\pm$ 131 \\ \hline
 0.41 & CH$_2 $ & 4.7 & 948 $\pm$ 54 \\ \hline
 0.41 & CR39 & 7.4 & 1285 $\pm$ 245 \\ \hline
 0.41 & Al & 27 & 1950 $\pm$ 126 \\ \hline
 0.30 & CH$_2 $ & 4.7 & 949 $\pm$ 61 \\ \hline
 0.30 & CR39 & 7.4 & 1174 $\pm$ 192 \\ \hline
 0.30 & Al & 27 & 2008 $\pm$ 144 \\ \hline
\end{tabular}
}
\end {center}
\caption {Total fragmentation cross sections, with statistical 
standard deviations, for Fe$^{26+}$ ions of different energies (col. 1) on 
different targets (col. 2).} 
\label{table:1}
\end{table}

 In Fig. \ref{fig:3} our data are compared with the semi-empirical 
formula (Bradt, H.L. and Peters, B., 1950) for nuclear cross sections 
(solid lines)

\begin{equation}
\sigma_{tot} = \pi r_0^2~ (A_P^{1/3}  + A_T^{1/3} -b_0)^2
\label{eq:2}
\end{equation}
where $r_0 = 1.31$ fm, $b_0 = 1.0$, $A_P$ and $A_T$ are the projectile and 
target mass numbers, respectively.

\begin{figure*}[ht]
\centering
 {\centering\resizebox*{!}{5.8cm}{\includegraphics{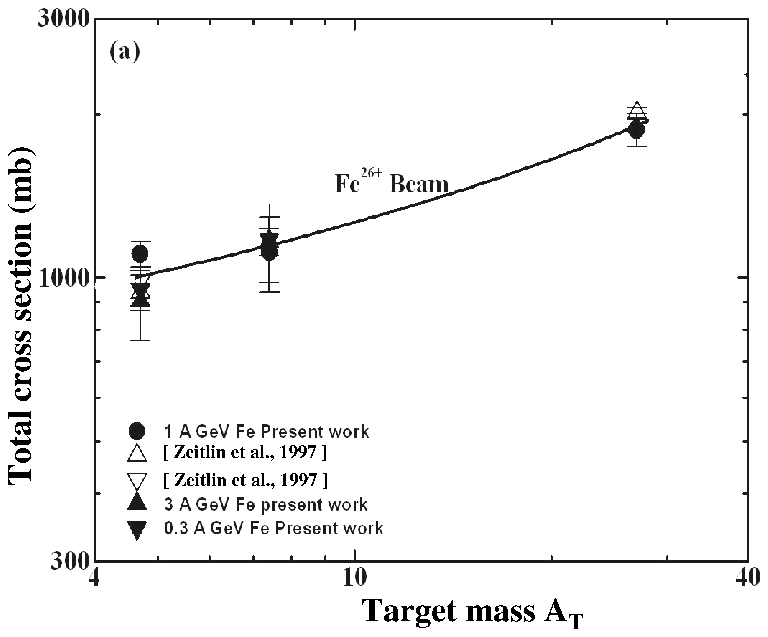}}}
\hspace{1cm}
 {\centering\resizebox*{!}{5.8cm}{\includegraphics{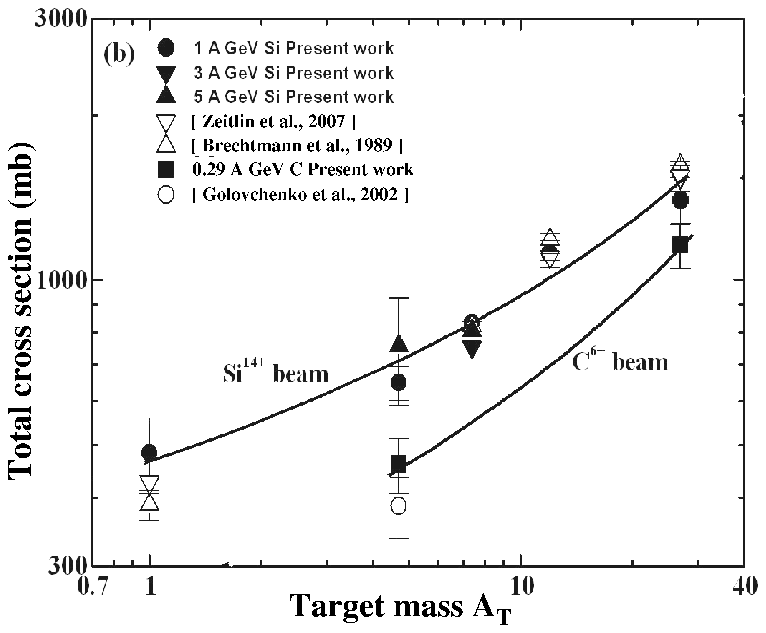}}\par}
\begin{quote}
\caption{\small Total fragmentation cross 
sections vs the target mass (a) for Fe$^{26+}$ ions and (b) for Si$^{14+}$ 
and C$^{6+}$ ions. 
The measured cross sections from the cited refs. 
are shown for comparison. The solid lines are from Eq. \ref{eq:2} 
corrected by the $\sigma_{EMD}$ term.} 
\label{fig:4}
\end{quote}
 \end{figure*}

Figs. \ref{fig:4}a,b show the total fragmentation cross sections vs target 
mass number $A_T$ for Fe$^{26+}$, Si$^{14+}$ and C$^{6+}$ beams of various 
energies. The solid lines are the predictions of Eq. \ref{eq:2}, to which 
we added the electromagnetic dissociation 
contribution, $\sigma_{EMD} = \alpha Z_T^{\delta}$, with 
$\alpha = 1.57$ fm$^2$ and $\delta = 
1.9$ (Dekhissi 2000). The total fragmentation cross 
sections increase with increasing target mass number. Part of the 
increase is due to the effect of electromagnetic dissociation. \par

The data from other authors (Brechtmann 1988; Brechtmann 1989; 
  Golovchenko 1999; Golovchenko 2002; Zeitlin 1997; Zeitlin 2007) are 
plotted for comparison
and show good agreement with our data, within the experimental uncertainties.

\begin{table}
\begin{center}
{\footnotesize
\begin{tabular}
{|c|c|c||c|c|c|}\hline
\multicolumn{3}{|c||}{\bf {\boldmath Si$^{14+}$} ions} & 
\multicolumn{3}{c|}{\bf {\boldmath C$^{6+}$} ions} \\ \hline
{\bf Energy} & {\bf Target} & {\bf {\boldmath $\sigma_{tot}$} (mb)} &
{\bf Energy} & {\bf Target} & {\bf {\boldmath $\sigma_{tot}$} (mb)} \\
{\bf (A GeV)} & & & {\bf (A GeV)} & &  \\ \hline
5 & CH$_2$ & 757 $\pm$ 168 & 0.29 & CH$_2$ & 460 $\pm$ 53 \\ \hline
3 & Al     & 1533 $\pm$ 133 & 0.29 & CR39 & 513 $\pm$ 52 \\ \hline
1 & CR39 & 1113 $\pm$ 176 & 0.29 & Al & 1155 $\pm$ 108 \\ \hline
1 & H & 483 $\pm$ 76 & & &  \\ \hline 
1 & CH$_2$ & 694 $\pm$ 70 & & & \\ \hline
1 & C & 1117 $\pm$ 62 & & & \\ \hline
1 & Al & 1397 $\pm$ 138 & & & \\ \hline

\end{tabular}
}
\end{center}
\caption {Total fragmentation cross sections, with
statistical standard deviations, for 
Si$^{14+}$ ions of different energies (col.1) on different targets 
(col. 2) and for 0.29 A GeV C$^{6+}$ ions on different targets 
(col. 5).} 
\label{table:2}
\end{table}

\section{Partial fragmentation charge changing cross sections}
If the thickness of the target is small compared to the mean free path of 
the fragments in that material, the partial fragmentation cross sections 
can be calculated using the simple relation
\begin{equation}
\sigma(Z_i, Z_f) \simeq \frac{1}{Kt} \frac{N_f}{N_i}
\end{equation}
where  $\sigma (Z_i, Z_f)$ is the partial fragmentation cross section of an 
ion $Z_i$  into the fragment $Z_f$, $K$ is the number of target nuclei 
per cm$^3$, $t$ is the thickness of the target, $N_i$  is the number of 
survived ions after the target and $N_f$ is the number of fragments produced 
with charge $Z_f$. This expression may be valid also for a thick 
target, assuming that the number of fragments before the target is zero. \par

For the Fe$^{26+}$ ions, we observed that fragments are present 
even before the 
targets. In this case the partial charge change cross 
sections have been computed via the relation

\begin{equation}
\sigma_{\Delta Z} = \frac{1}{Kt} \left( \frac{N^f_{out}}{N^p_s} - \frac{N^f_{in}}{N^p_{in}} \right)
\end{equation}
where $N_{in}^{f}$ and $N_{out}^f$ are the numbers of fragments of each 
charge before and after the target, and $N_{in}^{p}$ and $N_{s}^p$ 
are the numbers of incident and survived projectile ions. \par
The distributions, after the CH$_2$ targets, of the fragments for 1 A GeV 
Si$^{14+}$ and 1 A GeV 
Fe$^{26+}$ ions are shown in Figs. \ref{fig:2}a,b. The relative partial 
fragmentation cross sections for 
$\Delta Z = -1, -2, -3,~..,~ -18$ are given in Table \ref{table:3}. The quoted 
errors are statistical standard deviations; systematic uncertainties are 
estimated to be about $10\%$. A clear odd-even effect is visible in Fig. 
\ref{fig:2}: the cross sections for the $Z-$even fragments are generally 
larger than those for the $Z-$odd fragments close~by.

\section{Conclusions}
The total fragmentation cross sections for Fe$^{26+}$, Si$^{14+}$ and C$^{6+}$ 
ion beams of $0.3 \div 10$ A GeV energies on polyethylene, CR39 
and aluminum targets were measured using CR39 NTD's.

The total cross sections for all the targets and energies used in the 
present work do not show any observable energy 
dependence. There is a dependence on target mass; the
highest cross sections are observed for Al targets and this is mainly due to 
the contribution of electromagnetic dissociation. The present data of 
total fragmentation cross sections are in agreement with similar 
experimental data in the literature (Brechtmann 1988;
 Brechtmann 1989; Cecchini 2002; Dekhissi 2000; Flesch 2001; Golovchenko 
1999; Golovchenko 2002; Zeitlin 1997; Zeitlin 2007).

\begin{table}
\begin{center}
{\small
\begin{tabular}
{|c|c|c|}\hline
{\bf {\boldmath $\Delta Z$} } & {\bf 1 A GeV {\boldmath Fe$^{26+}$}} & 
{\bf 1 A GeV {\boldmath Si$^{14+}$}} \\ \hline
-1 & - & 293 $\pm$ 18 \\ \hline
-2 & 338 $\pm$ 11 & 177 $\pm$ 12 \\ \hline
-3 & 285 $\pm$ 11 & 123 $\pm$ 11 \\ \hline
-4 & 252 $\pm$ 10 & 122 $\pm$ 11 \\ \hline
-5 & 249 $\pm$ 10 & 62 $\pm$ 8 \\ \hline
-6 & 197 $\pm$ 9  & 117 $\pm$ 11 \\ \hline
-7 & 168 $\pm$ 8  & 83 $\pm$ 9 \\ \hline
-8 & 132 $\pm$ 7  & 90 $\pm$ 10 \\ \hline
-9 & 175 $\pm$ 8  & \\ \hline
-10 & 107 $\pm$ 7 & \\ \hline
-11 & 152 $\pm$ 6 & \\ \hline
-12 & 105 $\pm$ 8 & \\ \hline
-13 & 103 $\pm$ 6 & \\ \hline
-14 & 81 $\pm$ 6 & \\ \hline
-15 & 80 $\pm$ 6 & \\ \hline
-16 & 50 $\pm$ 4 & \\ \hline
-17 & 76 $\pm$ 5 & \\ \hline
-18 & 86 $\pm$ 6 & \\ \hline
\end{tabular}
}
\end {center}
\caption {Partial fragmentation charge changing cross sections, with
 statistical standard deviations,
 for 1 AGeV Si$^{14+}$ and Fe$^{26+}$ ions on the CH$_2$ targets.} 
\label{table:3}
\end{table}

The presence of well separated fragment peaks, see Fig. \ref{fig:2}, allowed
the determination of the partial fragmentation cross sections. On the 
average the partial cross sections decrease as the 
charge change $\Delta Z$ increases. The data in Fig. \ref{fig:2} and the 
partial cross sections in Table \ref{table:3} indicate a clear $Z$ odd-even 
effect. \par

The measured cross section data indicate that passive NTD's, specifically 
CR39, can be used effectively for studies of the total and partial charge 
changing cross sections, also in comparison with active detectors. 

\section*{Acknowledgments} 

We acknowledge the grants NRA-02-OBPR-02 and 15B517 for
the BIOSHIELD and BINFRA experiments, respectively. 

We thank the technical staff of BNL and HIMAC for their kind cooperation 
during the beam exposures and the contribution of our technical 
staff. We thank INFN and ICTP for providing fellowships and grants 
to non-Italian citizens.

\end{document}